\documentclass[
    aps,
    prd,
    amssymb,
    superscriptaddress,
    floatfix,
    nofootinbib,
    reprint
]{revtex4-2}

\usepackage{amsmath}
\usepackage{amsfonts}
\usepackage{bm}
\usepackage{braket}
\usepackage{graphicx}
\usepackage{booktabs}
\usepackage{multirow}
\usepackage{dcolumn}
\usepackage{enumitem}
\usepackage{lineno}
\usepackage[dvipsnames,x11names]{xcolor}
\usepackage[normalem]{ulem}
\usepackage{soul}
\usepackage{hyperref}
\hypersetup{
    colorlinks=true,
    citecolor=Green,
    urlcolor=Blue,
    linkcolor=Blue
}
\usepackage{comment}
\usepackage{mfirstuc}
\usepackage{orcidlink}

\usepackage{tikz}
\usepackage[compat=1.1.0]{tikz-feynman}
\tikzfeynmanset{arrow size=1}

\usepackage{caption}
\usepackage{float}

\usepackage[T1]{fontenc}
\usepackage{microtype}
\usepackage[vvarbb,smallerops,lining]{newtx}

\usepackage[compact]{titlesec}

\titleformat{\section}
    {\centering\small\bfseries\MakeUppercase}
    {\noindent\thesection.}
    {2pt}{}

\titleformat{\subsection}
    {\centering\small\bfseries}
    {\noindent\thesubsection.}
    {2pt}{}

\titlespacing\section{0pt}{18pt plus 1pt minus 1pt}{8pt plus 1pt minus 1pt}
\titlespacing\subsection{0pt}{12pt plus 1pt minus 1pt}{8pt plus 1pt minus 1pt}

\makeatletter

\newdimen\tov@rt
\newcommand{\tovgap}{2}

\newcommand{\tov@setrule}[1]{%
    \ifx#1\scriptscriptstyle
        \tov@rt=\fontdimen8\scriptscriptfont3
    \else\ifx#1\scriptstyle
        \tov@rt=\fontdimen8\scriptfont3
    \else
        \tov@rt=\fontdimen8\textfont3
    \fi\fi
}

\newcommand{\tov@build}[2]{%
    \tov@setrule{#1}
    \sbox0{$\m@th#1#2$}%
    \vbox{%
        \offinterlineskip
        \kern\tov@rt
        \hrule height\tov@rt
        \kern\tovgap\tov@rt
        \box0
    }
}

\newcommand{\tightoverline}[1]{\mathpalette\tov@build{#1}}

\makeatother

\newcommand{\olsi}[1]{\,\tightoverline{\!#1}}

\begin{document}

\preprint{APS/123-QED}

\title{$ \bar K^0 d $ Bound State and Correlation Function }% Force line breaks with \\

\author{M. Bayar \orcidlink{0000-0002-5914-0126}}
	\email{melahat.bayar@istanbul.edu.tr}
	\affiliation{Department of Physics, Faculty of Sciences, Istanbul University, 34134 Istanbul, Turkey}
	\affiliation{Departamento de F\'{i}sica Teórica and IFIC, Centro Mixto Universidad de Valencia-CSIC, Institutos de Investigaci\'{o}n de Paterna, Aptdo. 22085, E-46071 Valencia, Spain}
	
\author{P. Encarnación \orcidlink{0009-0005-0749-3885}}
	\email{pablo.encarnacion@ific.uv.es}
	\affiliation{Departamento de F\'{i}sica Teórica and IFIC, Centro Mixto Universidad de Valencia-CSIC, Institutos de Investigaci\'{o}n de Paterna, Aptdo. 22085, E-46071 Valencia, Spain}
 
\author{A. Feijoo \orcidlink{0000-0002-8580-802X}}
	\email{Eduardo.Feijoo@ific.uv.es}
	\affiliation{Departamento de F\'{i}sica Teórica and IFIC, Centro Mixto Universidad de Valencia-CSIC, Institutos de Investigaci\'{o}n de Paterna, Aptdo. 22085, E-46071 Valencia, Spain}
 
\author{E. Oset \orcidlink{0000-0002-4462-7919}}
	\email{Eulogio.Oset@ific.uv.es}
	\affiliation{Departamento de F\'{i}sica Teórica and IFIC, Centro Mixto Universidad de Valencia-CSIC, Institutos de Investigaci\'{o}n de Paterna, Aptdo. 22085, E-46071 Valencia, Spain}
	 \affiliation{Department of Physics, Guangxi Normal University, Guilin 541004, China}
   
\date{\today}% It is always \today, today,

\begin{abstract}
We study the $\bar{K}^0d$ system using a new formulation of the fixed-center approximation that preserves elastic unitarity. The resulting scattering amplitude predicts a bound state below threshold, which can be interpreted as a $\Lambda(1380)N$ configuration complementary to the $K^-pp$ state associated with the higher $\Lambda(1405)$ pole. The interaction is attractive, with scattering parameters of the same order as recent $K^-d$ determinations. We further calculate the $\bar{K}^0d$ correlation function, which exhibits a characteristic structure associated with the bound state and provides a Coulomb-free observable that can be tested in future ALICE measurements. Results obtained with two different $\bar{K}N$ models are very similar, indicating the robustness of the predictions against the choice of the underlying interaction.
\end{abstract}

%\keywords{Suggested keywords}%Use showkeys class option if keyword                   %display desired
\maketitle

%\tableofcontents

\section{Introduction}
\label{sec:introduction}

Historically, kaonic atoms have provided a sensitive probe of low-energy strong interactions, as the large kaon mass leads to compact atomic orbits, bringing the antikaon close to the nucleus and inducing measurable strong-interaction shifts and broadenings of the low-lying atomic levels~\cite{Zmeskal:2008zz}. In particular, the ground-state shift and width of kaonic hydrogen and deuterium are related to the complex $\olsi{K}N$ scattering lengths, whose combined analysis allows the determination of the isoscalar and isovector components. This bears directly on the $K^-n$ scattering amplitude, which is purely $I=1$ and for which chirally motivated benchmark models show substantial differences, as illustrated in Fig.~1 of Ref.~\cite{Cieply:2020ftt}.

These scattering lengths and the corresponding subthreshold amplitude are commonly obtained from coupled-channel chiral models. The $\olsi{K}N$ interaction has attracted considerable theoretical interest for decades, owing to its attractive character and the presence of several resonances around the $\olsi{K}N$ threshold~\cite{dalitz,thomas,fink,rubin}. In this context, Unitarized Chiral Perturbation Theory (UChPT) has proven to be a powerful framework for describing low-energy meson-baryon interactions in the $S=-1$ sector, successfully reproducing experimental data and dynamically generating bound states and resonances~\cite{KSW,KWW,Oset:1997it,OM,Garcia-Recio:2002yxy,BNW}. Of particular interest is the $\Lambda(1405)$, originally proposed in the 1950s~\cite{L1405}, whose properties have been interpreted in terms of two poles generated by coupled-channel meson-baryon dynamics~\cite{OM,2pole,PRL,Nieves:2024dcz,Mai:2020ltx,BaryonScatteringBaSc:2023zvt,Zhuang:2024udv}. Only recently has this interpretation been incorporated into the PDG compilation \cite{ParticleDataGroup:2020ssz}. The availability of increasingly precise experimental data has enabled tighter constraints on chiral unitary models, showing that while the Weinberg--Tomozawa (WT) interaction dominates the classical $\olsi{K}N$ channels, higher-order contributions become essential when incorporating additional reactions and global data sets across different strangeness sectors, providing important constraints on the corresponding low-energy constants (LECs)~\cite{IHW,HJ_rev,Cieply:2011nq,GO,Mai:2014xna,Feijoo:2015yja,
Ramos:2016odk,Lu:2022hwm,Feijoo:2018den}.

Kaonic hydrogen has been measured with increasing precision~\cite{Iwasaki:1997wf,DEAR:2005fdl,DEAR:2005ueb,SIDD,Bazzi:2026zmc}, whereas the experimental determination of kaonic deuterium has remained challenging because of its extremely low X-ray yield~\cite{SIDDHARTA:2013ftj}. The SIDDHARTA-2 measurement~\cite{Bazzi:2026gri} now provides the first direct experimental information on kaonic deuterium, adding an independent constraint to the low-energy
$\olsi{K}N$ interaction and complementing the existing kaonic-hydrogen and scattering data. The result is therefore particularly relevant to the poorly constrained isovector interaction. Prior to this measurement, theoretical predictions for the $K^-d$ scattering length and the corresponding $1s$ level shift and width varied substantially with the theoretical approach, ranging from fixed-center approximation (FCA), through Faddeev calculations, to more sophisticated three-body treatments~\cite{Barrett:1999cw,Deloff:1999gc,Kamalov:2000iy,Bahaoui:2002yc,Borasoy:2005ie,Gal:2006cw,Doring:2011xc,Shevchenko:2011ce,Shevchenko:2012np,Revai:2012fx,Mizutani:2012gy,Mai:2014uma,Revai:2016muw,Hoshino:2017mty,Liu:2020foc,Duerinck:2025pbj}.

In recent years, experimental efforts have established femtoscopy as a promising tool for extracting information on hadron--hadron interactions. In particular, the measured $K^-p$ correlation function (CF)~\cite{ALICE:2019gcn} provided the first experimental evidence for the opening of the $\olsi{K}^0 n$ channel, demonstrating the sensitivity of femtoscopic correlations to the underlying coupled-channel dynamics~\cite{Kamiya:2019uiw}. While an initial analysis reported tensions with scattering data~\cite{ALICE:2022yyh}, a subsequent theoretical analysis~\cite{Encarnacion:2024jge} found compatibility with the femtoscopic data using leading- and next-to-leading-order UChPT models constrained by scattering data~\cite{2pole,Feijoo:2018den}. A further step was taken in Ref.~\cite{Xie:2026hpp}, where this problem was addressed within a fully off-shell covariant UChPT framework. These developments provided the basis for the recent $K^-d$ CF measurement~\cite{ALICE:2026pxr}, offering the first experimental access to the low-energy $K^-d$ interaction. In parallel, a study within the FCA and impulse approximations~\cite{Ramos:2025ibe} found good agreement with the ALICE data.

Furthermore, the $K^-d$ system is closely connected to the study of few-body antikaon--nuclear systems. Its interaction has also been discussed in the context of possible quasi-bound or resonant $\olsi{K}NN$ states \cite{Oset:2012gi}. The properties of such few-body systems depend sensitively on the energy dependence of the underlying $\olsi{K}N$ interaction~\cite{Dote:2008in}, making the $K^-d$ system particularly relevant for constraining its subthreshold behavior.

In this study, we investigate the $\olsi{K}^0 d$ scattering amplitude and CF using a new formulation of the FCA, in which elastic unitarity is explicitly preserved. The neutral system is chosen to eliminate Coulomb effects, allowing a clean probe of the strong interaction. We determine the low-energy scattering parameters and the subthreshold amplitude, and study the resulting femtoscopic CF using different $\olsi{K}N$ interaction models to assess theoretical uncertainties.

\section{Formalism}
\label{sec:Formalism}

We closely follow the formalism developed for the study of the $p f_1(1285)$ interaction in Ref.~\cite{Encarnacion:2026zas}. The $\olsi K^0$ has isospin $I=1/2$, while the deuteron is an isoscalar ($I=0$) state, $|d\rangle=(|pn\rangle-|np\rangle)/\sqrt{2}$. Thus, as in the $p f_1(1285)$ system, the total isospin of the scattering system is $I=1/2$. In Ref.~\cite{Encarnacion:2026zas}, the $f_1(1285)$ was treated as a $K^*\olsi K-\olsi K^*K$ molecular state, whereas here the cluster is the deuteron. We use a $\olsi K^0$ rather than a $K^-$ to avoid Coulomb effects at small relative momentum and thus isolate the strong $\olsi Kd$
interaction. This also distinguishes the present study from previous calculations of the $K^-d$ scattering length discussed above.
\begin{figure}
    $
    \begin{array}{ccccccc}
        &
        \begin{tikzpicture}[scale=0.9, baseline={(current bounding box.center)}]
            \begin{feynman}
                % Define vertices
                \vertex (name1) at (0,-0.5) {$\frac{1}{\sqrt{2}}\left( pn - np \right)$};
                \vertex (name2) at (0,3.25) {};
                \vertex (iL) at (-0.25,0) {};
                \vertex (iR) at (0.25,0) {};
                \vertex (iN) at (-0.75,0.1) {$\olsi{K}^0$};
                \vertex (fL) at (-0.25,3) {};
                \vertex (fR) at (0.25,3) {};
                \vertex (fN) at (-0.75,2.9) {$\olsi{K}^0$};
                \vertex (v1) [dot, style={fill,scale=0.01}] at (-0.25,1.5) {};
                \vertex (v2) [dot, style={fill,scale=0.01}] at (0.25,1.5) {};
                
                % First diagram: Initial interaction of N with K1
                \diagram* {
                    (name1) -- [boson, opacity=0.0] (name2),
                    (iN) -- [dashed, fermion] (v1) -- [dashed, fermion] (fN),
                    (iL) -- [fermion] (v1) -- [fermion] (fL),
                    (iR) -- [fermion] (v2) -- [fermion] (fR)
                };
                \filldraw[fill=black!20, draw=black]
                    (0,0.3) ellipse (0.5 and 0.1);
        
                \filldraw[fill=black!20, draw=black]
                    (0,2.7) ellipse (0.5 and 0.1);
                
            \end{feynman}
        \end{tikzpicture}
        & + &
        \begin{tikzpicture}[scale=0.9, baseline={(current bounding box.center)}]
            \begin{feynman}
                % Define vertices
                \vertex (name1) at (0,-0.5) {$\frac{1}{\sqrt{2}}\left( pn - np \right)$};
                \vertex (name2) at (0,3.25) {};
                \vertex (iL) at (-0.25,0) {};
                \vertex (iR) at (0.25,0) {};
                \vertex (iN) at (-0.75,0.1) {$\olsi{K}^0$};
                \vertex (fL) at (-0.25,3) {};
                \vertex (fR) at (0.25,3) {};
                \vertex (fN) at (0.75,2.9) {$\olsi{K}^0$};
                \vertex (v1) [dot, style={fill,scale=0.01}] at (-0.25,1.5) {};
                \vertex (v2) [dot, style={fill,scale=0.01}] at (0.25,1.5) {};
                
                % First diagram: Initial interaction of N with K1
                \diagram* {
                    (name1) -- [boson, opacity=0.0] (name2),
                    (iN) -- [dashed, fermion] (v1) -- [scalar] (v2) -- [dashed, fermion] (fN),
                    (iL) -- [fermion] (v1) -- [fermion] (fL),
                    (iR) -- [fermion] (v2) -- [fermion] (fR)
                };
                \filldraw[fill=black!20, draw=black]
                    (0,0.3) ellipse (0.5 and 0.1);
        
                \filldraw[fill=black!20, draw=black]
                    (0,2.7) ellipse (0.5 and 0.1);
                
            \end{feynman}
        \end{tikzpicture}
        & + &
        \begin{tikzpicture}[scale=0.9, baseline={(current bounding box.center)}]
            \begin{feynman}
                % Define vertices
                \vertex (name1) at (0,-0.5) {$\frac{1}{\sqrt{2}}\left( pn - np \right)$};
                \vertex (name2) at (0,3.25) {};
                \vertex (iL) at (-0.25,0) {};
                \vertex (iR) at (0.25,0) {};
                \vertex (iN) at (-0.75,0.1) {$\olsi{K}^0$};
                \vertex (fL) at (-0.25,3) {};
                \vertex (fR) at (0.25,3) {};
                \vertex (fN) at (-0.75,2.9) {$\olsi{K}^0$};
                \vertex (v1) [dot, style={fill,scale=0.01}] at (-0.25,1) {};
                \vertex (v2) [dot, style={fill,scale=0.01}] at (0.25,1.5) {};
                \vertex (v3) [dot, style={fill,scale=0.01}] at (-0.25,2) {};
                
                % First diagram: Initial interaction of N with K1
                \diagram* {
                    (name1) -- [boson, opacity=0.0] (name2),
                    (iN) -- [dashed, fermion] (v1) -- [scalar] (v2) -- [scalar] (v3) -- [dashed, fermion] (fN),
                    (iL) -- [fermion] (v1) -- [fermion] (v3) -- [fermion] (fL),
                    (iR) -- [fermion] (v2) -- [fermion] (fR)
                };
                \filldraw[fill=black!20, draw=black]
                    (0,0.3) ellipse (0.5 and 0.1);
        
                \filldraw[fill=black!20, draw=black]
                    (0,2.7) ellipse (0.5 and 0.1);
                
            \end{feynman}
        \end{tikzpicture}
    & +\ \dots \\
        + &
        \begin{tikzpicture}[scale=0.9, baseline={(current bounding box.center)}]
            \begin{feynman}
                % Define vertices
                \vertex (name1) at (0,-0.5) {$\frac{1}{\sqrt{2}}\left( pn - np \right)$};
                \vertex (name2) at (0,3.25) {};
                \vertex (iL) at (-0.25,0) {};
                \vertex (iR) at (0.25,0) {};
                \vertex (iN) at (0.75,0.1) {$\olsi{K}^0$};
                \vertex (fL) at (-0.25,3) {};
                \vertex (fR) at (0.25,3) {};
                \vertex (fN) at (0.75,2.9) {$\olsi{K}^0$};
                \vertex (v1) [dot, style={fill,scale=0.01}] at (-0.25,1.5) {};
                \vertex (v2) [dot, style={fill,scale=0.01}] at (0.25,1.5) {};

                % First diagram: Initial interaction of N with K1
                \diagram* {
                    (name1) -- [boson, opacity=0.0] (name2),
                    (iN) -- [dashed, fermion] (v2) -- [dashed, fermion] (fN),
                    (iL) -- [fermion] (v1) -- [fermion] (fL),
                    (iR) -- [fermion] (v2) -- [fermion] (fR)
                };
                \filldraw[fill=black!20, draw=black]
                    (0,0.3) ellipse (0.5 and 0.1);
        
                \filldraw[fill=black!20, draw=black]
                    (0,2.7) ellipse (0.5 and 0.1);
            \end{feynman}
        \end{tikzpicture}
        & + &
        \begin{tikzpicture}[scale=0.9, baseline={(current bounding box.center)}]
            \begin{feynman}
                % Define vertices
                \vertex (name1) at (0,-0.5) {$\frac{1}{\sqrt{2}}\left( pn - np \right) $};
                \vertex (name2) at (0,3.25) {};
                \vertex (iL) at (-0.25,0) {};
                \vertex (iR) at (0.25,0) {};
                \vertex (iN) at (0.75,0.1) {$\olsi{K}^0$};
                \vertex (fL) at (-0.25,3) {};
                \vertex (fR) at (0.25,3) {};
                \vertex (fN) at (-0.75,2.9) {$\olsi{K}^0$};
                \vertex (v1) [dot, style={fill,scale=0.01}] at (-0.25,1.5) {};
                \vertex (v2) [dot, style={fill,scale=0.01}] at (0.25,1.5) {};

                % Second diagram: Interaction propagates inside the system
                \diagram* {
                    (name1) -- [boson, opacity=0.0] (name2),
                    (iN) -- [dashed, fermion] (v2) -- [scalar] (v1) -- [dashed, fermion] (fN),
                    (iL) -- [fermion] (v1) -- [fermion] (fL),
                    (iR) -- [fermion] (v2) -- [fermion] (fR)
                };
                \filldraw[fill=black!20, draw=black]
                    (0,0.3) ellipse (0.5 and 0.1);
        
                \filldraw[fill=black!20, draw=black]
                    (0,2.7) ellipse (0.5 and 0.1);
            \end{feynman}
        \end{tikzpicture}
        & + &
        \begin{tikzpicture}[scale=0.9, baseline={(current bounding box.center)}]
            \begin{feynman}
                % Define vertices
                \vertex (name1) at (0,-0.5) {$\frac{1}{\sqrt{2}}\left( pn - np \right)$};
                \vertex (name2) at (0,3.25) {};
                \vertex (iL) at (-0.25,0) {};
                \vertex (iR) at (0.25,0) {};
                \vertex (iN) at (0.75,0.1) {$\olsi{K}^0$};
                \vertex (fL) at (-0.25,3) {};
                \vertex (fR) at (0.25,3) {};
                \vertex (fN) at (0.75,2.9) {$\olsi{K}^0$};
                \vertex (v1) [dot, style={fill,scale=0.01}] at (-0.25,1.5) {};
                \vertex (v2) [dot, style={fill,scale=0.01}] at (0.25,1) {};
                \vertex (v3) [dot, style={fill,scale=0.01}] at (0.25,2) {};

                % Third diagram: Extended rescattering
                \diagram* {
                    (name1) -- [boson, opacity=0.0] (name2),
                    (iN) -- [dashed, fermion] (v2) -- [scalar] (v1) -- [scalar] (v3) -- [dashed, fermion] (fN),
                    (iL) -- [fermion] (v1) -- [fermion] (fL),
                    (iR) -- [fermion] (v2) -- [fermion] (v3) -- [fermion] (fR)
                };
                \filldraw[fill=black!20, draw=black]
                    (0,0.3) ellipse (0.5 and 0.1);
        
                \filldraw[fill=black!20, draw=black]
                    (0,2.7) ellipse (0.5 and 0.1);
            \end{feynman}
        \end{tikzpicture}
    & +\ \dots
    \end{array}
    $
    \caption{Diagrams in the FCA for $ \olsi K^0 d $ scattering.}
    \label{fig:diagramsFCA}
\end{figure}
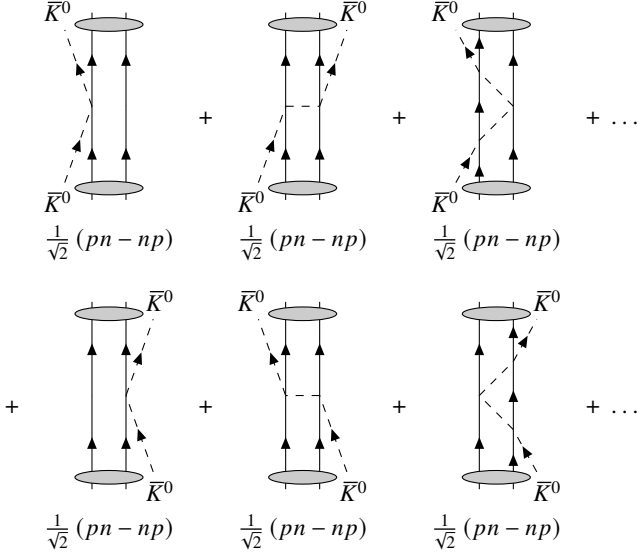
We employ a new formulation of the FCA, where elastic unitarity is implemented~\cite{Ikeno:2025bsx}. This feature is important for a consistent treatment of the amplitude at and above threshold. This method provides a link with the framework used to study elastic scattering of external particles from nuclei, through the introduction of an optical potential and the solution of the corresponding Lippmann--Schwinger equation. The new formulation has been used to study the interaction of external particles with molecular resonances, including $n\olsi{D}_{s0}^*(2317)$~\cite{Ikeno:2025bsx}, $n\olsi{D}_{s1}(2460)$ and $n\olsi{D}_{s1}(2536)$~\cite{Agatao:2025ckp}, $Kf_1(1285)$~\cite{Jia:2026dJpl}, $\pi f_1(1285)$ and $\eta f_1(1285)$~\cite{Jia:2026ewk}, $KX(3872)$ system~\cite{Song:2026spz}, as well as the very recent study of the $\pi T_{cc}$ system~\cite{Brandao:2026bmr}. These developments illustrate the applicability of the method to systems involving molecular resonances, with the $pf_1(1285)$ study of Ref.~\cite{Encarnacion:2026zas} providing a direct femtoscopic benchmark, as its predictions were subsequently supported by the ALICE measurement~\cite{Serksnyte2026presentation}. This successful result motivates the present study of the $\olsi K^0d$ system.

\subsection{$\olsi{K}^0 d$ scattering amplitude}

The standard FCA describes the interaction of an external particle, here the $\olsi K^0$, with a cluster of two particles, here the deuteron, which remains unchanged during the interaction. This is analogous to elastic scattering of an external particle from a nucleus, where the nucleus remains intact throughout the collision. Starting from the diagrams shown in Fig.~\ref{fig:diagramsFCA}, the FCA constructs the partition functions $T_{ij}$, with $ i,~j=1,~2 $, where $i$ indicates that the $\olsi K^0$ interacts first with particle $i$ of the deuteron, $\left( pn - np \right)/\sqrt{2}$, and finishes with the $\olsi K^0$ emitted from particle $j$. These partition functions are easily found and read\footnote{ We follow Mandl and Shaw \cite{mandl_shaw_2010} normalization for the meson and baryon fields, that is, $ 1/ \sqrt{2\omega} $ factor for mesons and $  \sqrt{2M/2E} $ for baryons. In addition we also use the field normalization $ \sqrt{ 2 M_{d} / 2 E_{d} } $ for the deuteron and consider $ \sqrt{2 M_{i}/2 E_{i} } \approx 1 $. As a consequence, there is no need of defining the $ \tilde{t}_{i} $  amplitudes that one has in the case of mesons in the cluster. Explicit derivation can be seen in Ref. \cite{Bayar:2011qj}.}
\begin{equation}
\begin{aligned}
T_{11} &= \frac{t_1}{1-t_1t_2G_0^2}, \qquad
T_{12}=T_{21} = \frac{t_1t_2G_0}{1-t_1t_2G_0^2}, \\
T_{22} &= \frac{t_2}{1-t_1t_2G_0^2}.
\end{aligned}
\label{eq:TT}
\end{equation}
where $t_1$ is the scattering matrix for the interaction of the $\olsi K^0$ with particle $1$ of the cluster, and $t_2$ the corresponding matrix for particle $2$. Taking into account that the cluster is the $I=0$ deuteron state, $d=(pn-np)/\sqrt{2}$, one readily finds~\cite{Roca:2010tf,Xiao:2011rc,Ikeno:2022jbb,Bayar:2023itf}
\begin{equation}
   t_1  \equiv   t_2 = \frac{3}{4} t_{\olsi{K} N }^{I=1} + \frac{1}{4} t_{\olsi{K} N }^{I=0}
    \label{eq:tKd}
\end{equation}
where we take the $ s $-wave, spin independent, $ t_{\olsi{K} N } $, from \cite{Oset:1997it}. 

The $G_0$ function in Eq.~(\ref{eq:TT}) is the $ \olsi K^0 $ propagator through the cluster, modulated by the deuteron wave function, which is written as 
\begin{eqnarray}
    G_0(\sqrt{s}) \!=\!\! \!\! & \displaystyle\int & \!\!\!\!\! \frac{d^3 q}{(2\pi)^3}  \frac{1 }{2\omega_{ \olsi K^0}(q)}  \frac{2M_d }{2E_{d}(q)}\frac{F_C(q)}{\sqrt{s} \!-\! \omega_{ \olsi K^0}(q) \!-\! E_{d}(q) \!+\! i\epsilon} \nonumber \\
    &&\times
     \theta(q_{max}^{(1)}-q_1^*)\theta(q_{max}^{(2)}-q_2^*)
    \label{eq:G0}
\end{eqnarray}
with  $ M_d $ being the deuteron mass, and $\omega_{ \olsi K^0}(q)$ , $E_{d}(q)$ representing the energies of the $ \olsi K^0  $ and the deuteron, respectively. In addition $ F_C(q) $ is the deuteron form factor which is taken directly from \cite{Machleidt:2000ge,Bayar:2011qj}, normalized to $F_{C}(0)=1$ for the $ s $ wave function of the deuteron.
\begin{figure}
    \centering
        \begin{tikzpicture}[scale=0.9, baseline={(current bounding box.center)}, inner sep=0pt, outer sep=0pt]]
            \begin{feynman}
                % Define vertices
                \vertex (iL) at (-0.25,0) {};
                \vertex (iR) at (0.25,0) {};
                \vertex (iN) at (-0.8,0) {};
                \vertex (fL) at (-0.25,3) {};
                \vertex (fR) at (0.25,3) {};
                \vertex (fN) at (-0.8,3) {};
                \vertex (ghost1) at (-0.8,0) {};
                \vertex (ghost2) at (0.8,3) {};
                
                \vertex (v11) [dot, style={fill,scale=0.01}] at (-0.25,0.6) {};
                \vertex (v12) [dot, style={fill,scale=0.01}] at (-0.25,1.2) {};
                
                \vertex (v21) [dot, style={fill,scale=0.01}] at (-0.25,1.8) {};
                \vertex (v22) [dot, style={fill,scale=0.01}] at (-0.25,2.4) {};
                
                \vertex (v111) [dot, style={fill,scale=0.01}] at (0.25,0.75) {};
                \vertex (v112) [dot, style={fill,scale=0.01}] at (-0.25,0.9) {};
                \vertex (v113) [dot, style={fill,scale=0.01}] at (0.25,1.05) {};
                
                \vertex (v211) [dot, style={fill,scale=0.01}] at (0.25,1.95) {};
                \vertex (v212) [dot, style={fill,scale=0.01}] at (-0.25,2.1) {};
                \vertex (v213) [dot, style={fill,scale=0.01}] at (0.25,2.25) {};
                
                % First diagram: Initial interaction of N with K1
                \diagram* {
                    (iN) -- [dashed, fermion, arrow size=0.8pt, dash pattern=on 2pt off 1pt] (v11) 
                    -- [dashed, fermion, arrow size=0.8pt, dash pattern=on 2pt off 1pt] (v111) 
                    -- [dashed, fermion, arrow size=0.8pt, dash pattern=on 2pt off 1pt] (v112) 
                    -- [dashed, fermion, arrow size=0.8pt, dash pattern=on 2pt off 1pt] (v113) 
                    -- [dashed, fermion, arrow size=0.8pt, dash pattern=on 2pt off 1pt] (v12) 
                    -- [dashed, fermion, arrow size=0.8pt, dash pattern=on 2pt off 1pt, half left] (v21) 
                    -- [dashed, fermion, arrow size=0.8pt, dash pattern=on 2pt off 1pt] (v211) 
                    -- [dashed, fermion, arrow size=0.8pt, dash pattern=on 2pt off 1pt] (v212) 
                    -- [dashed, fermion, arrow size=0.8pt, dash pattern=on 2pt off 1pt] (v213) 
                    -- [dashed, fermion, arrow size=0.8pt, dash pattern=on 2pt off 1pt] (v22) 
                    -- [dashed, fermion, arrow size=0.8pt, dash pattern=on 2pt off 1pt] (fN) ,
                    (iL) -- (v1) --  (fL),
                    (iR) --  (v2) --  (fR),
                    (ghost1) -- [boson, opacity=0.0] (ghost2),
                };
                \filldraw[fill=black!20, draw=black]
                    (0,0.3) ellipse (0.4 and 0.1);
                \filldraw[fill=black!20, draw=black]
                    (0,1.5) ellipse (0.4 and 0.1);
                \filldraw[fill=black!20, draw=black]
                    (0,2.7) ellipse (0.4 and 0.1);
                
            \end{feynman}
        \end{tikzpicture} 
        \hspace{0pt} $+$ \hspace{0pt} 
        \begin{tikzpicture}[scale=0.9, baseline={(current bounding box.center)}, inner sep=0pt, outer sep=0pt]]
            \begin{feynman}
                % Define vertices
                \vertex (iL) at (-0.25,0) {};
                \vertex (iR) at (0.25,0) {};
                \vertex (iN) at (-0.8,0) {};
                \vertex (fL) at (-0.25,3) {};
                \vertex (fR) at (0.25,3) {};
                \vertex (fN) at (-0.8,3) {};
                \vertex (ghost1) at (-0.8,0) {};
                \vertex (ghost2) at (0.8,3) {};
                
                \vertex (v11) [dot, style={fill,scale=0.01}] at (-0.25,0.6) {};
                \vertex (v12) [dot, style={fill,scale=0.01}] at (0.25,1.2) {};
                
                \vertex (v21) [dot, style={fill,scale=0.01}] at (0.25,1.8) {};
                \vertex (v22) [dot, style={fill,scale=0.01}] at (-0.25,2.4) {};
                
                \vertex (v111) [dot, style={fill,scale=0.01}] at (0.25,0.8) {};
                \vertex (v112) [dot, style={fill,scale=0.01}] at (-0.25,1) {};
                
                \vertex (v211) [dot, style={fill,scale=0.01}] at (-0.25,2) {};
                \vertex (v212) [dot, style={fill,scale=0.01}] at (0.25,2.2) {};
                
                % First diagram: Initial interaction of N with K1
                \diagram* {
                    (iN) -- [dashed, fermion, arrow size=0.8pt, dash pattern=on 2pt off 1pt] (v11) 
                    -- [dashed, fermion, arrow size=0.8pt, dash pattern=on 2pt off 1pt] (v111) 
                    -- [dashed, fermion, arrow size=0.8pt, dash pattern=on 2pt off 1pt] (v112) 
                    -- [dashed, fermion, arrow size=0.8pt, dash pattern=on 2pt off 1pt] (v12) 
                    -- [dashed, fermion, arrow size=0.8pt, dash pattern=on 2pt off 1pt, half right] (v21) 
                    -- [dashed, fermion, arrow size=0.8pt, dash pattern=on 2pt off 1pt] (v211) 
                    -- [dashed, fermion, arrow size=0.8pt, dash pattern=on 2pt off 1pt] (v212) 
                    -- [dashed, fermion, arrow size=0.8pt, dash pattern=on 2pt off 1pt] (v22) 
                    -- [dashed, fermion, arrow size=0.8pt, dash pattern=on 2pt off 1pt] (fN) ,
                    (iL) -- (v1) --  (fL),
                    (iR) --  (v2) --  (fR),
                    (ghost1) -- [boson, opacity=0.0] (ghost2),
                };
                \filldraw[fill=black!20, draw=black]
                    (0,0.3) ellipse (0.4 and 0.1);
                \filldraw[fill=black!20, draw=black]
                    (0,1.5) ellipse (0.4 and 0.1);
                \filldraw[fill=black!20, draw=black]
                    (0,2.7) ellipse (0.4 and 0.1);
                
            \end{feynman}
        \end{tikzpicture}
        \hspace{0pt} $+$ \hspace{0pt} 
        \begin{tikzpicture}[scale=0.9, baseline={(current bounding box.center)}, inner sep=0pt, outer sep=0pt]]
            \begin{feynman}
                % Define vertices
                \vertex (iL) at (-0.25,0) {};
                \vertex (iR) at (0.25,0) {};
                \vertex (iN) at (-0.8,0) {};
                \vertex (fL) at (-0.25,3) {};
                \vertex (fR) at (0.25,3) {};
                \vertex (fN) at (0.8,3) {};
                \vertex (ghost1) at (-0.8,0) {};
                \vertex (ghost2) at (0.8,3) {};
                
                \vertex (v11) [dot, style={fill,scale=0.01}] at (-0.25,0.6) {};
                \vertex (v12) [dot, style={fill,scale=0.01}] at (-0.25,1.2) {};
                
                \vertex (v21) [dot, style={fill,scale=0.01}] at (-0.25,1.8) {};
                \vertex (v22) [dot, style={fill,scale=0.01}] at (0.25,2.4) {};
                
                \vertex (v111) [dot, style={fill,scale=0.01}] at (0.25,0.75) {};
                \vertex (v112) [dot, style={fill,scale=0.01}] at (-0.25,0.9) {};
                \vertex (v113) [dot, style={fill,scale=0.01}] at (0.25,1.05) {};
                
                \vertex (v211) [dot, style={fill,scale=0.01}] at (0.25,2) {};
                \vertex (v212) [dot, style={fill,scale=0.01}] at (-0.25,2.2) {};
                
                % First diagram: Initial interaction of N with K1
                \diagram* {
                    (iN) -- [dashed, fermion, arrow size=0.8pt, dash pattern=on 2pt off 1pt] (v11) 
                    -- [dashed, fermion, arrow size=0.8pt, dash pattern=on 2pt off 1pt] (v111) 
                    -- [dashed, fermion, arrow size=0.8pt, dash pattern=on 2pt off 1pt] (v112) 
                    -- [dashed, fermion, arrow size=0.8pt, dash pattern=on 2pt off 1pt] (v113) 
                    -- [dashed, fermion, arrow size=0.8pt, dash pattern=on 2pt off 1pt] (v12) 
                    -- [dashed, fermion, arrow size=0.8pt, dash pattern=on 2pt off 1pt, half left] (v21) 
                    -- [dashed, fermion, arrow size=0.8pt, dash pattern=on 2pt off 1pt] (v211) 
                    -- [dashed, fermion, arrow size=0.8pt, dash pattern=on 2pt off 1pt] (v212) 
                    -- [dashed, fermion, arrow size=0.8pt, dash pattern=on 2pt off 1pt] (v22) 
                    -- [dashed, fermion, arrow size=0.8pt, dash pattern=on 2pt off 1pt] (fN) ,
                    (iL) -- (v1) --  (fL),
                    (iR) --  (v2) --  (fR),
                    (ghost1) -- [boson, opacity=0.0] (ghost2),
                };
                \filldraw[fill=black!20, draw=black]
                    (0,0.3) ellipse (0.4 and 0.1);
                \filldraw[fill=black!20, draw=black]
                    (0,1.5) ellipse (0.4 and 0.1);
                \filldraw[fill=black!20, draw=black]
                    (0,2.7) ellipse (0.4 and 0.1);
                
            \end{feynman}
        \end{tikzpicture}
        \hspace{0pt} $+$ \hspace{0pt} 
        \begin{tikzpicture}[scale=0.9, baseline={(current bounding box.center)}, inner sep=0pt, outer sep=0pt]]
            \begin{feynman}
                % Define vertices
                \vertex (iL) at (-0.25,0) {};
                \vertex (iR) at (0.25,0) {};
                \vertex (iN) at (-0.8,0) {};
                \vertex (fL) at (-0.25,3) {};
                \vertex (fR) at (0.25,3) {};
                \vertex (fN) at (0.8,3) {};
                \vertex (ghost1) at (-0.8,0) {};
                \vertex (ghost2) at (0.8,3) {};
                
                \vertex (v11) [dot, style={fill,scale=0.01}] at (-0.25,0.6) {};
                \vertex (v12) [dot, style={fill,scale=0.01}] at (0.25,1.2) {};
                
                \vertex (v21) [dot, style={fill,scale=0.01}] at (0.25,1.8) {};
                \vertex (v22) [dot, style={fill,scale=0.01}] at (0.25,2.4) {};
                
                \vertex (v111) [dot, style={fill,scale=0.01}] at (0.25,0.8) {};
                \vertex (v112) [dot, style={fill,scale=0.01}] at (-0.25,1) {};
                
                \vertex (v211) [dot, style={fill,scale=0.01}] at (-0.25,1.95) {};
                \vertex (v212) [dot, style={fill,scale=0.01}] at (0.25,2.1) {};
                \vertex (v213) [dot, style={fill,scale=0.01}] at (-0.25,2.25) {};
                
                % First diagram: Initial interaction of N with K1
                \diagram* {
                    (iN) -- [dashed, fermion, arrow size=0.8pt, dash pattern=on 2pt off 1pt] (v11) 
                    -- [dashed, fermion, arrow size=0.8pt, dash pattern=on 2pt off 1pt] (v111) 
                    -- [dashed, fermion, arrow size=0.8pt, dash pattern=on 2pt off 1pt] (v112) 
                    -- [dashed, fermion, arrow size=0.8pt, dash pattern=on 2pt off 1pt] (v12) 
                    -- [dashed, fermion, arrow size=0.8pt, dash pattern=on 2pt off 1pt, half right] (v21) 
                    -- [dashed, fermion, arrow size=0.8pt, dash pattern=on 2pt off 1pt] (v211) 
                    -- [dashed, fermion, arrow size=0.8pt, dash pattern=on 2pt off 1pt] (v212) 
                    -- [dashed, fermion, arrow size=0.8pt, dash pattern=on 2pt off 1pt] (v213) 
                    -- [dashed, fermion, arrow size=0.8pt, dash pattern=on 2pt off 1pt] (v22) 
                    -- [dashed, fermion, arrow size=0.8pt, dash pattern=on 2pt off 1pt] (fN) ,
                    (iL) -- (v1) --  (fL),
                    (iR) --  (v2) --  (fR),
                    (ghost1) -- [boson, opacity=0.0] (ghost2),
                };
                \filldraw[fill=black!20, draw=black]
                    (0,0.3) ellipse (0.4 and 0.1);
                \filldraw[fill=black!20, draw=black]
                    (0,1.5) ellipse (0.4 and 0.1);
                \filldraw[fill=black!20, draw=black]
                    (0,2.7) ellipse (0.4 and 0.1);
                
            \end{feynman}
        \end{tikzpicture} 
        \\ [5pt]
        \begin{tikzpicture}[scale=0.9, baseline={(current bounding box.center)}, inner sep=0pt, outer sep=0pt]]
            \begin{feynman}
                % Define vertices
                \vertex (iL) at (-0.25,0) {};
                \vertex (iR) at (0.25,0) {};
                \vertex (iN) at (0.8,0) {};
                \vertex (fL) at (-0.25,3) {};
                \vertex (fR) at (0.25,3) {};
                \vertex (fN) at (-0.8,3) {};
                \vertex (ghost1) at (-0.8,0) {};
                \vertex (ghost2) at (0.8,3) {};
                
                \vertex (v11) [dot, style={fill,scale=0.01}] at (0.25,0.6) {};
                \vertex (v12) [dot, style={fill,scale=0.01}] at (-0.25,1.2) {};
                
                \vertex (v21) [dot, style={fill,scale=0.01}] at (-0.25,1.8) {};
                \vertex (v22) [dot, style={fill,scale=0.01}] at (-0.25,2.4) {};
                
                \vertex (v111) [dot, style={fill,scale=0.01}] at (-0.25,0.8) {};
                \vertex (v112) [dot, style={fill,scale=0.01}] at (0.25,1) {};
                
                \vertex (v211) [dot, style={fill,scale=0.01}] at (0.25,1.95) {};
                \vertex (v212) [dot, style={fill,scale=0.01}] at (-0.25,2.1) {};
                \vertex (v213) [dot, style={fill,scale=0.01}] at (0.25,2.25) {};
                
                % First diagram: Initial interaction of N with K1
                \diagram* {
                    (iN) -- [dashed, fermion, arrow size=0.8pt, dash pattern=on 2pt off 1pt] (v11) 
                    -- [dashed, fermion, arrow size=0.8pt, dash pattern=on 2pt off 1pt] (v111) 
                    -- [dashed, fermion, arrow size=0.8pt, dash pattern=on 2pt off 1pt] (v112) 
                    -- [dashed, fermion, arrow size=0.8pt, dash pattern=on 2pt off 1pt] (v12) 
                    -- [dashed, fermion, arrow size=0.8pt, dash pattern=on 2pt off 1pt, half left] (v21) 
                    -- [dashed, fermion, arrow size=0.8pt, dash pattern=on 2pt off 1pt] (v211) 
                    -- [dashed, fermion, arrow size=0.8pt, dash pattern=on 2pt off 1pt] (v212) 
                    -- [dashed, fermion, arrow size=0.8pt, dash pattern=on 2pt off 1pt] (v213) 
                    -- [dashed, fermion, arrow size=0.8pt, dash pattern=on 2pt off 1pt] (v22) 
                    -- [dashed, fermion, arrow size=0.8pt, dash pattern=on 2pt off 1pt] (fN) ,
                    (iL) -- (v1) --  (fL),
                    (iR) --  (v2) --  (fR),
                    (ghost1) -- [boson, opacity=0.0] (ghost2),
                };
                \filldraw[fill=black!20, draw=black]
                    (0,0.3) ellipse (0.4 and 0.1);
                \filldraw[fill=black!20, draw=black]
                    (0,1.5) ellipse (0.4 and 0.1);
                \filldraw[fill=black!20, draw=black]
                    (0,2.7) ellipse (0.4 and 0.1);
                
            \end{feynman}
        \end{tikzpicture}
        \hspace{0pt} $+$ \hspace{0pt} 
        \begin{tikzpicture}[scale=0.9, baseline={(current bounding box.center)}, inner sep=0pt, outer sep=0pt]]
            \begin{feynman}
                % Define vertices
                \vertex (iL) at (-0.25,0) {};
                \vertex (iR) at (0.25,0) {};
                \vertex (iN) at (0.8,0) {};
                \vertex (fL) at (-0.25,3) {};
                \vertex (fR) at (0.25,3) {};
                \vertex (fN) at (-0.8,3) {};
                \vertex (ghost1) at (-0.8,0) {};
                \vertex (ghost2) at (0.8,3) {};
                
                \vertex (v11) [dot, style={fill,scale=0.01}] at (0.25,0.6) {};
                \vertex (v12) [dot, style={fill,scale=0.01}] at (0.25,1.2) {};
                
                \vertex (v21) [dot, style={fill,scale=0.01}] at (0.25,1.8) {};
                \vertex (v22) [dot, style={fill,scale=0.01}] at (-0.25,2.4) {};
                
                \vertex (v111) [dot, style={fill,scale=0.01}] at (-0.25,0.75) {};
                \vertex (v112) [dot, style={fill,scale=0.01}] at (0.25,0.9) {};
                \vertex (v113) [dot, style={fill,scale=0.01}] at (-0.25,1.05) {};
                
                \vertex (v211) [dot, style={fill,scale=0.01}] at (-0.25,2) {};
                \vertex (v212) [dot, style={fill,scale=0.01}] at (0.25,2.2) {};
                
                % First diagram: Initial interaction of N with K1
                \diagram* {
                    (iN) -- [dashed, fermion, arrow size=0.8pt, dash pattern=on 2pt off 1pt] (v11) 
                    -- [dashed, fermion, arrow size=0.8pt, dash pattern=on 2pt off 1pt] (v111) 
                    -- [dashed, fermion, arrow size=0.8pt, dash pattern=on 2pt off 1pt] (v112) 
                    -- [dashed, fermion, arrow size=0.8pt, dash pattern=on 2pt off 1pt] (v113) 
                    -- [dashed, fermion, arrow size=0.8pt, dash pattern=on 2pt off 1pt] (v12) 
                    -- [dashed, fermion, arrow size=0.8pt, dash pattern=on 2pt off 1pt, half right] (v21) 
                    -- [dashed, fermion, arrow size=0.8pt, dash pattern=on 2pt off 1pt] (v211) 
                    -- [dashed, fermion, arrow size=0.8pt, dash pattern=on 2pt off 1pt] (v212) 
                    -- [dashed, fermion, arrow size=0.8pt, dash pattern=on 2pt off 1pt] (v22) 
                    -- [dashed, fermion, arrow size=0.8pt, dash pattern=on 2pt off 1pt] (fN) ,
                    (iL) -- (v1) --  (fL),
                    (iR) --  (v2) --  (fR),
                    (ghost1) -- [boson, opacity=0.0] (ghost2),
                };
                \filldraw[fill=black!20, draw=black]
                    (0,0.3) ellipse (0.4 and 0.1);
                \filldraw[fill=black!20, draw=black]
                    (0,1.5) ellipse (0.4 and 0.1);
                \filldraw[fill=black!20, draw=black]
                    (0,2.7) ellipse (0.4 and 0.1);
                
            \end{feynman}
        \end{tikzpicture}
        \hspace{0pt} $+$ \hspace{0pt} 
        \begin{tikzpicture}[scale=0.9, baseline={(current bounding box.center)}, inner sep=0pt, outer sep=0pt]]
            \begin{feynman}
                % Define vertices
                \vertex (iL) at (-0.25,0) {};
                \vertex (iR) at (0.25,0) {};
                \vertex (iN) at (0.8,0) {};
                \vertex (fL) at (-0.25,3) {};
                \vertex (fR) at (0.25,3) {};
                \vertex (fN) at (0.8,3) {};
                \vertex (ghost1) at (-0.8,0) {};
                \vertex (ghost2) at (0.8,3) {};
                
                \vertex (v11) [dot, style={fill,scale=0.01}] at (0.25,0.6) {};
                \vertex (v12) [dot, style={fill,scale=0.01}] at (-0.25,1.2) {};
                
                \vertex (v21) [dot, style={fill,scale=0.01}] at (-0.25,1.8) {};
                \vertex (v22) [dot, style={fill,scale=0.01}] at (0.25,2.4) {};
                
                \vertex (v111) [dot, style={fill,scale=0.01}] at (-0.25,0.8) {};
                \vertex (v112) [dot, style={fill,scale=0.01}] at (0.25,1) {};
                
                \vertex (v211) [dot, style={fill,scale=0.01}] at (0.25,2) {};
                \vertex (v212) [dot, style={fill,scale=0.01}] at (-0.25,2.2) {};
                
                % First diagram: Initial interaction of N with K1
                \diagram* {
                    (iN) -- [dashed, fermion, arrow size=0.8pt, dash pattern=on 2pt off 1pt] (v11) 
                    -- [dashed, fermion, arrow size=0.8pt, dash pattern=on 2pt off 1pt] (v111) 
                    -- [dashed, fermion, arrow size=0.8pt, dash pattern=on 2pt off 1pt] (v112) 
                    -- [dashed, fermion, arrow size=0.8pt, dash pattern=on 2pt off 1pt] (v12) 
                    -- [dashed, fermion, arrow size=0.8pt, dash pattern=on 2pt off 1pt, half left] (v21) 
                    -- [dashed, fermion, arrow size=0.8pt, dash pattern=on 2pt off 1pt] (v211) 
                    -- [dashed, fermion, arrow size=0.8pt, dash pattern=on 2pt off 1pt] (v212) 
                    -- [dashed, fermion, arrow size=0.8pt, dash pattern=on 2pt off 1pt] (v22) 
                    -- [dashed, fermion, arrow size=0.8pt, dash pattern=on 2pt off 1pt] (fN) ,
                    (iL) -- (v1) --  (fL),
                    (iR) --  (v2) --  (fR),
                    (ghost1) -- [boson, opacity=0.0] (ghost2),
                };
                \filldraw[fill=black!20, draw=black]
                    (0,0.3) ellipse (0.4 and 0.1);
                \filldraw[fill=black!20, draw=black]
                    (0,1.5) ellipse (0.4 and 0.1);
                \filldraw[fill=black!20, draw=black]
                    (0,2.7) ellipse (0.4 and 0.1);
                
            \end{feynman}
        \end{tikzpicture}
        \hspace{0pt} $+$ \hspace{0pt} 
        \begin{tikzpicture}[scale=0.9, baseline={(current bounding box.center)}, inner sep=0pt, outer sep=0pt]
            \begin{feynman}
                % Define vertices
                \vertex (iL) at (-0.25,0) {};
                \vertex (iR) at (0.25,0) {};
                \vertex (iN) at (0.8,0) {};
                \vertex (fL) at (-0.25,3) {};
                \vertex (fR) at (0.25,3) {};
                \vertex (fN) at (0.8,3) {};
                \vertex (ghost1) at (-0.8,0) {};
                \vertex (ghost2) at (0.8,3) {};
                
                \vertex (v11) [dot, style={fill,scale=0.01}] at (0.25,0.6) {};
                \vertex (v12) [dot, style={fill,scale=0.01}] at (0.25,1.2) {};
                
                \vertex (v21) [dot, style={fill,scale=0.01}] at (0.25,1.8) {};
                \vertex (v22) [dot, style={fill,scale=0.01}] at (0.25,2.4) {};
                
                \vertex (v111) [dot, style={fill,scale=0.01}] at (-0.25,0.75) {};
                \vertex (v112) [dot, style={fill,scale=0.01}] at (0.25,0.9) {};
                \vertex (v113) [dot, style={fill,scale=0.01}] at (-0.25,1.05) {};
                
                \vertex (v211) [dot, style={fill,scale=0.01}] at (-0.25,1.95) {};
                \vertex (v212) [dot, style={fill,scale=0.01}] at (0.25,2.1) {};
                \vertex (v213) [dot, style={fill,scale=0.01}] at (-0.25,2.25) {};
                
                % First diagram: Initial interaction of N with K1
                \diagram* {
                    (iN) -- [dashed, fermion, arrow size=0.8pt, dash pattern=on 2pt off 1pt] (v11) 
                    -- [dashed, fermion, arrow size=0.8pt, dash pattern=on 2pt off 1pt] (v111) 
                    -- [dashed, fermion, arrow size=0.8pt, dash pattern=on 2pt off 1pt] (v112) 
                    -- [dashed, fermion, arrow size=0.8pt, dash pattern=on 2pt off 1pt] (v113) 
                    -- [dashed, fermion, arrow size=0.8pt, dash pattern=on 2pt off 1pt] (v12) 
                    -- [dashed, fermion, arrow size=0.8pt, dash pattern=on 2pt off 1pt, half right] (v21) 
                    -- [dashed, fermion, arrow size=0.8pt, dash pattern=on 2pt off 1pt] (v211) 
                    -- [dashed, fermion, arrow size=0.8pt, dash pattern=on 2pt off 1pt] (v212) 
                    -- [dashed, fermion, arrow size=0.8pt, dash pattern=on 2pt off 1pt] (v213) 
                    -- [dashed, fermion, arrow size=0.8pt, dash pattern=on 2pt off 1pt] (v22) 
                    -- [dashed, fermion, arrow size=0.8pt, dash pattern=on 2pt off 1pt] (fN) ,
                    (iL) -- (v1) --  (fL),
                    (iR) --  (v2) --  (fR),
                    (ghost1) -- [boson, opacity=0.0] (ghost2),
                };
                \filldraw[fill=black!20, draw=black]
                    (0,0.3) ellipse (0.4 and 0.1);
                \filldraw[fill=black!20, draw=black]
                    (0,1.5) ellipse (0.4 and 0.1);
                \filldraw[fill=black!20, draw=black]
                    (0,2.7) ellipse (0.4 and 0.1);
                
            \end{feynman}
        \end{tikzpicture}
    \caption{Diagrams considered in the unitarization of the $\olsi{K}^{0} d$ amplitude.}
    \label{fig:diagramsunitarity}
\end{figure}
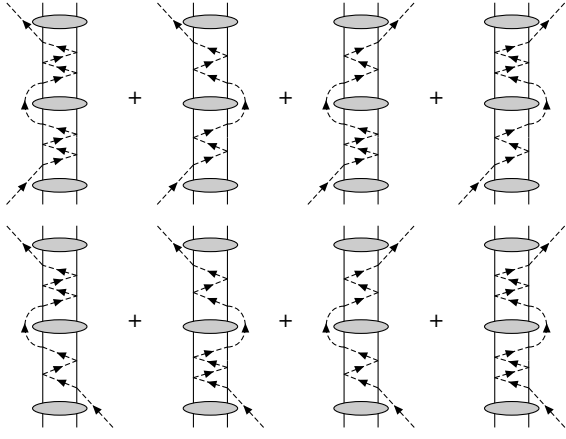
The $\theta(q_{\max}^{(i)}-q_i^*)$ functions in Eq.~\eqref{eq:G0} originate from the momentum cutoff used in the $\olsi{K} N$ scattering matrices of~\cite{Oset:1997it}. These matrices are expressed as $T =\left[ 1-V G \right]^{-1} V$, where $V$ is the transition potential between the coupled channels---namely, $\olsi{K} N$, $\pi \Sigma$, $\pi \Lambda$, $\eta \Sigma$, $\eta \Lambda$, and $K \Xi$---while $G$ represents the meson-baryon loop function regularized by a cutoff $q_{\max}$. As shown in~\cite{Gamermann:2009uq}, this is equivalent to using a separable potential of the type $V \theta (q_{\max}-|\vec{q}|) \theta (q_{\max}-|\vec{q'}|)$, which implies that the $t$-matrix inherits the same $\theta$ functions: $t (\vec{q} , \vec{q}')=t~ \theta (q_{\max}-|\vec{q}|) \theta (q_{\max}-|\vec{q'}|)$. We adopt here the value $q_{\max}=630$~MeV used in~\cite{Oset:1997it}, setting $q^{(1)}_{\max} = q^{(2)}_{\max}$. On the other hand, the magnitudes $q_i^*$ stand for the modulus of the $\olsi{K}^0$ momentum in the $\olsi{K}^0 N$ rest frame
\begin{eqnarray}
    \vec q_i^* = \vec q \left( 1 - \frac{1}{2} \frac{M_{\olsi K^0}}{M_{\olsi K^0}+M_i} \right), ~~i=1,~2 \,.
\end{eqnarray}

As done in~\cite{Ikeno:2025bsx}, the partition functions $T_{ij}$ are considered as an optical potential that is iterated in the Lippmann-Schwinger series, as shown in Fig.~\ref{fig:diagramsunitarity} for just the second-order iteration. Resumming these diagrams and their successive iterations yields
\begin{equation}
     T' = [1 - T G_C]^{-1} T \, ,
\end{equation}
where $T$ denotes the $T_{ij}$ matrix of Eq.~\eqref{eq:TT}, while $G_C$ accounts for the propagation of the $\olsi{K}^{0}$ meson together with the cluster and is given by $G_C = \operatorname{diag}\left(G_C^{(1)}, G_C^{(2)}\right)$, with
\begin{eqnarray}
    G_C^{(i)}(\sqrt{s}) \!=\!\! \!\! & \displaystyle\int & \!\!\!\!\! \frac{d^3 q}{(2\pi)^3}  \frac{1 }{2\omega_{ \olsi K^0}(q)}  \frac{2M_d }{2E_{d}(q)}\frac{[F_C^{(i)}(q)]^2}{\sqrt{s} \!-\! \omega_{ \olsi K^0}(q) \!-\! E_{d}(q) \!+\! i\epsilon} \nonumber \\
    &&\times
     \theta(q_{\max}^{(i)}-q_i^*) \, ,
    \label{eq:GC}
\end{eqnarray}
where, following Ref.~\cite{Yamagata-Sekihara:2010kpd}, we take
\begin{equation}
    F_C^{(1)}(q) = F_C\left(\frac{M_N}{M_N+M_N}q\right) = F_C^{(2)}(q) \, .
\end{equation}
The final total amplitude, $T^{\rm tot} = \sum_{i,j} T_{ij}'$, is given by~\cite{Agatao:2025ckp}
\begin{equation}
    T^{\rm tot} =\frac{ t_1+ t_2 + (2G_0-G_C^{(1)}-G_C^{(2)})t_1 t_2}{1-G_C^{(1)} t_1-G_C^{(2)} t_2-(G_0^2-G_C^{(1)}G_C^{(2)}) t_1 t_2} \, .
    \label{eq:TotalT}
\end{equation}

\subsection{Scattering  Length and Effective Range}

The amplitude obtained in Eq.~\eqref{eq:TotalT} can be related to the ordinary scattering amplitude used in Quantum Mechanics via
\begin{equation}
    -\frac{2 M_{d}}{8\pi\sqrt{s}}T^{\rm tot} \equiv f^{\rm QM} \simeq \frac{1}{-\frac{1}{a}+\frac{1}{2}r_0q_{\rm cm}^2 - i q_{\rm cm}} \, ,
    \label{eq:EffRanExp}
\end{equation}
with $q_{\rm cm}$ being the $\olsi{K}^{0}$ momentum in the $\olsi{K}^{0} d$ rest frame. From this expression, one immediately obtains
\begin{equation}
    a = \frac{2 M_{d}}{8\pi\sqrt{s}}T^{\rm tot}  \bigg|_{\rm th} \\
    \label{eq:a0}
\end{equation}
\begin{equation}
    r_0 = \frac{1}{\mu}\left[ \frac{\partial}{\partial\sqrt{s}} \left( -\frac{8\pi\sqrt{s}}{2M_d}(T^{\rm tot})^{-1}+iq_{cm} \right) \right]_{\rm th}
    \label{eq:r0}
\end{equation}
where the subscript ``$\rm th$'' indicates the $\olsi{K}^{0} d$ threshold.

In Eq.~\eqref{eq:EffRanExp}, $a$ and $r_0$ depend on the dynamics of the specific problem, whereas the $- i q_{\rm cm}$ term is universal and ensures the unitarity of the scattering matrix. Indeed, it was shown in Ref.~\cite{Agatao:2025ckp} that Eq.~\eqref{eq:TotalT} exactly satisfies the unitarity condition.

The final aspect required to complete the formalism is to specify the argument $s_{i}$ of the $t_{\olsi{K} N} (s_{i})$ matrices. We follow the standard procedure of distributing the binding energy of the cluster among its constituents proportionally to their masses (in this case, equally shared), which yields
\begin{eqnarray}
    s_1 = m_{\olsi{K}^{0}}^{2}+( M_N \xi )^2 + 2\xi M_N q^0 \equiv s_2 \, ,
    \label{eq:s1}
\end{eqnarray}
with $M_N$ being the nucleon mass, and $q^0$ representing the $\olsi{K}^{0}$ energy in the deuteron rest frame, $q^0=(s-m_{\olsi{K}^{0}}^2-M_d^2)/2M_d$, and $\xi = M_d/2 M_N$.

\subsection{Correlation function}

In addition, we calculate the $\olsi{K}^0 d$ CF, providing predictions that can be tested in future measurements by the ALICE Collaboration. The CF is evaluated as
\begin{equation}
\begin{split}
C_{\olsi{K}^{0} d}(p) &= 1+4\pi \int_0^\infty \!\! dr\,r^2 S_{12}(r) \\
&\quad\times \Big\{ |j_0(pr)+TG|^2-j_0^2(pr)\Big\} \, ,
\end{split}
\end{equation}
with
\begin{equation}
    TG = T^{\rm tot} \frac{1}{2}\Big( G_1(\sqrt{s},r) + G_2(\sqrt{s}, r) \Big) \, ,
    \label{eq:TG}
\end{equation}
where $S_{12}(r)$ is the source function, for which a conventional Gaussian form is assumed:
\begin{equation}
    S_{12}(r) = \frac{1}{(4\pi R^2)^{3/2}} e^{-r^2/4R^2} \, ,
\end{equation}
and
\begin{equation}
\begin{split}
    G_i(\sqrt{s},r) &= \!\! \int \!\!\! \frac{d^3 q}{(2\pi)^3} \frac{1}{2\omega_{\olsi{K}^0}(q)} \frac{2M_d}{2E_{d}(q)} \\
    &\quad\times \frac{F_C^{(i)}(q) j_0(qr)}{\sqrt{s} \!-\! \omega_{\olsi{K}^0}(q) \!-\! E_{d}(q) \!+\! i\epsilon} \theta(q_{\max}^{(i)} \!-\! q_i^*) \, ,
\end{split}
\end{equation}
which implies that $G_1(\sqrt{s},r) = G_2(\sqrt{s},r)$.

\section{Results}
\subsection{ $\olsi{K}^0 d$ scattering amplitude } 
To assess the theoretical uncertainties of our predictions, we consider two different $\olsi{K}N$ amplitudes. The first, from~\cite{Oset:1997it}, is based solely on the lowest-order WT interaction and will be referred to as the Oset--Ramos model. The second, from~\cite{Feijoo:2018den}, includes higher-order contributions from the chiral Lagrangian up to next-to-leading order (NLO), with the corresponding LECs fitted to a large set of experimental data; we will refer to this as the BCN model. Throughout the results, we also provide uncertainty estimates associated to each theoretical model. For the Oset-Ramos model we consider a variation in the cutoff that defines the two-body amplitudes $q_{\max}=630\pm50$~MeV. For the BCN model we propagate the uncertainties of the LECs reported in Ref.~\cite{Feijoo:2018den}. 

\begin{figure}
    \centering
    \includegraphics[width=\linewidth]{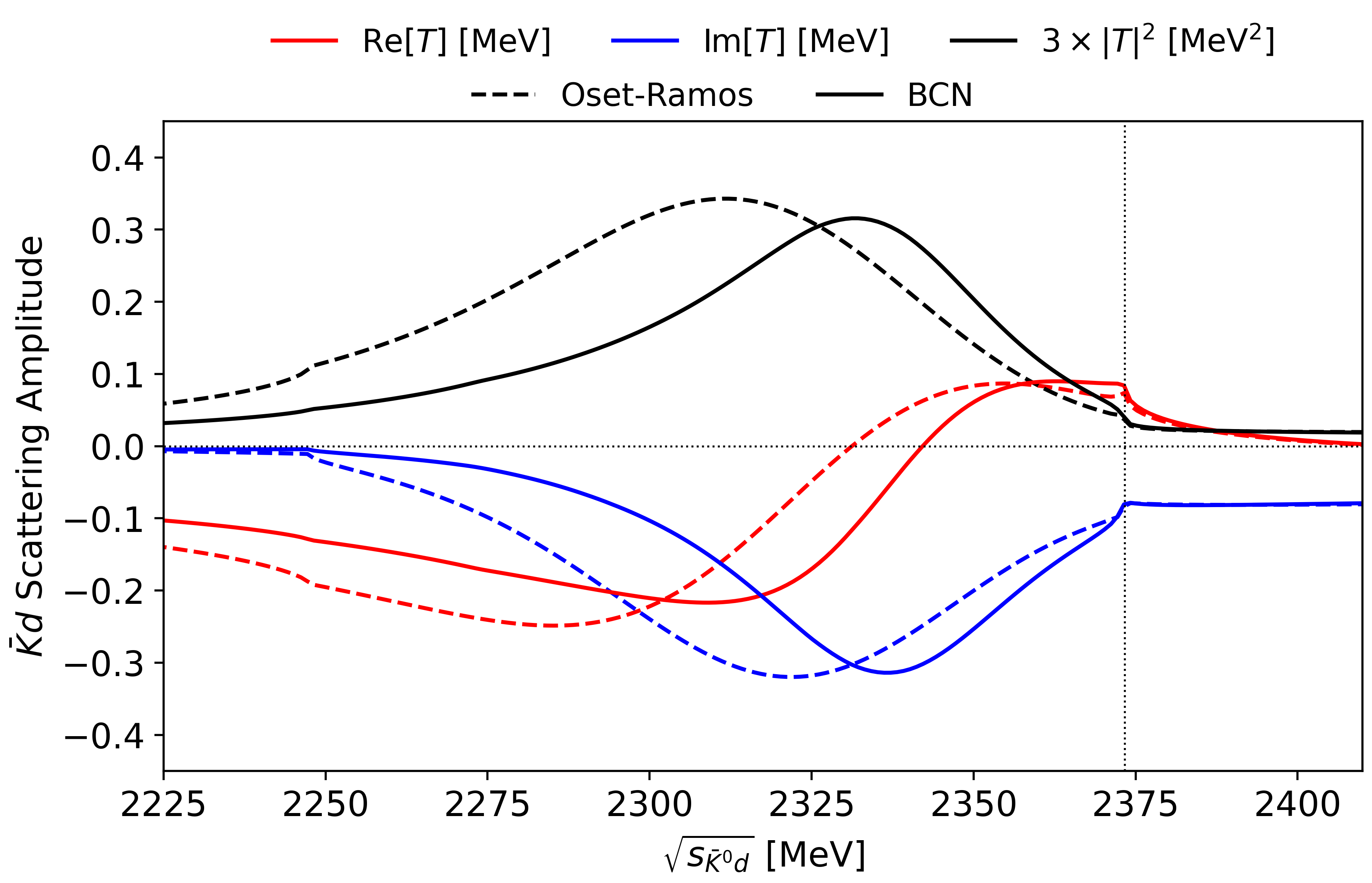}
    \captionsetup{skip=2pt}
    \caption{Results for $T^{\rm tot}$ obtained with the two theoretical models: Oset--Ramos~\cite{Oset:1997it} (dashed lines) and BCN~\cite{Feijoo:2018den} (solid lines). The real and imaginary parts of $T^{\rm tot}$ are represented by red and blue lines, respectively, while $|T^{\rm tot}|^2$ is shown multiplied by a factor of 3 for visualization purposes.}
    \label{fig:Kd_Tmatrix}
\end{figure}

In Fig.~\ref{fig:Kd_Tmatrix}, we show the results for the $\olsi{K}^{0}d$ scattering amplitude. In this figure, we observe a resonant structure below the $\olsi{K}^{0}d$ threshold, corresponding to a bound state of these components. The state, however, has a relatively large width, arising from the $\olsi{K}N\to\pi\Sigma$ decay channel. The resonance signal is particularly clean: ${\rm Im}[T^{\rm tot}]$ is negative and exhibits a clear peak in $|{\rm Im}[T^{\rm tot}]|$, while ${\rm Re}[T^{\rm tot}]$ changes sign from negative to positive, passing through zero close to the position of the $|{\rm Im}[T^{\rm tot}]|$ peak. We also observe a Breit--Wigner-like structure in $|T^{\rm tot}|^2$. Interestingly, the positions of the $|T^{\rm tot}|^2$ peaks obtained with the two models differ by about $20$ MeV, with the Oset--Ramos model predicting a more deeply bound state than the BCN one. The corresponding masses, binding energies, and widths are summarized in Table~\ref{tab:massandWidthK0d}.
\begin{table}[H]
    \centering
    \begin{tabular}{cccc}  
        \toprule 
        Model & Mass [MeV] &~ Binding [MeV]  & ~Width [MeV] \\ 
        \midrule 
         Oset--Ramos \cite{Oset:1997it} & $ 2314^{+24}_{-31} $ & $58^{+31}_{-24}$  & $ 77^{+3}_{-21}$ \\[1mm]
        BCN \cite{Feijoo:2018den} & $ 2333^{+5}_{-13} $ & $39^{+13}_{-5}$  & $ 54^{+23}_{-7}$ \\        
        \bottomrule 
    \end{tabular}
    \caption{ Mass, binding energy, and width of the predicted $\olsi{K}^{0}d$ bound state. }
    \label{tab:massandWidthK0d}
\end{table}
The prediction of a $\olsi{K}^{0}d$ bound state is not unexpected, given the extensive theoretical literature on the $K^-pp$ bound state~\citep{Ikeda:2007nz,Shevchenko:2006xy,Shevchenko:2007zz,Dote:2008in,Dote:2008hw,Ikeda:2008ub,Nogami:1963xqa,Yamazaki:2002uh,Bayar:2011qj,Kezerashvili:2016ztc,Filikhin:2020sar,Marri:2026qbj,Deng:2007zza,Yamazaki:2007cs,Revai:2014twa,Dote:2017wkk}, as well as on the $K^-d$ bound state~\citep{Oset:2012gi,Shevchenko:2020kpk}. In particular, recent experimental searches have provided evidence for the $K^-pp$ bound state through the study of the $^3{\rm He}(K^-,\Lambda p)n$ reaction~\cite{J-PARCE15:2018zys,J-PARCE15:2020gbh}. A theoretical analysis of this reaction, emphasizing the relevant role played by the $K^-pp$ bound state, was presented in~\cite{Sekihara:2016vyd}. More recently, the LEPS collaboration has also reported evidence for this state in the $\gamma d\to K^0\Lambda p$ reaction~\cite{LEPS2Solenoid:2026obe}. The latter experiment reports a binding energy of about $16$ MeV and a width of about $55$ MeV, compared with a binding energy of about $42$ MeV and a width of about $100$ MeV reported in the $K^-$-induced experiment~\cite{J-PARCE15:2018zys,J-PARCE15:2020gbh}. The discrepancy between the two experimental determinations likely points to larger uncertainties than those currently quoted. Taken together, the observation of a signal in two independent reactions provides strong support for the existence of the $K^-pp$ bound state and, by analogy, also supports the existence of the $\olsi{K}^{0}d$ bound state studied here, consistently with the important role of $\olsi{K}NN$ dynamics in few-body systems.

To interpret these results, we note that the resonance peak in Fig.~\ref{fig:Kd_Tmatrix} appears at $\sqrt{s}\simeq2314$~MeV for the Oset--Ramos model and at $\sqrt{s}\simeq2333$~MeV for the BCN model. The corresponding values of $\sqrt{s_1}$, obtained from Eq.~(\ref{eq:s1}), are $\sqrt{s_1}\simeq1386$~MeV and $\sqrt{s_1}\simeq1402$~MeV, respectively. These values lie very close to the energies of the lower $\Lambda(1405)$ poles obtained within the corresponding models. This lower pole is characterized by a stronger coupling to the $\pi\Sigma$ channels. In the Oset--Ramos model, its position is found at $1390$~MeV~\cite{2pole}, while in the BCN model it lies within the range $[1397,1435]$~MeV~\cite{Feijoo:2018den}. This suggests that the state obtained here can be interpreted as a $\Lambda(1380)N$ system.\footnote{Here, $\Lambda(1380)$ and $\Lambda(1420)$ are used as generic notation for the lower and higher $\Lambda^*$ poles, respectively, associated with the $\Lambda(1405)$ resonance.}

This picture is complementary to that of the $K^-pp$ bound state, which is associated with the higher $\Lambda(1405)$ pole and can thus be viewed as a $\Lambda(1420)N$ system. The $\olsi{K}^{0}d$ system therefore provides access to complementary information on the $\olsi{K}NN$ dynamics, in particular through the correlation function discussed below, whose low-momentum behavior is sensitive to the scattering parameters and to the presence of the predicted bound state. A future measurement of this correlation function would thus provide an independent test of our prediction. Complementary information could also be obtained from photoproduction reactions such as $\gamma d\to K^0\Sigma^0p$, where structures in the $M_{\Sigma p}$ invariant-mass distribution could provide access to the $\olsi{K}^{0}pn$ dynamics and offer an additional probe of the bound state predicted here.

\subsection{ Scattering  Length and Effective Range } 
\begin{table*}
    \centering
    \begin{tabular}{ccc}  
        \toprule 
          Theory ( $\olsi{K}^{0} d $)&~   $ a $ [fm] & ~$ r_{0} $ [fm] \\ 
        \midrule 
         Oset--Ramos   & $ 0.92^{+0.08}_{-0.03} - 0.98^{+0.05}_{-0.01} i$ & $ 2.9^{+1.4}_{-0.7} + 0.03^{+0.38}_{-0.82} i $   \\[2mm]
        BCN  & $ 1.04^{+0.19}_{-0.11} - 0.94^{+0.14}_{-0.03} i$ & $ 3.0^{+0.4}_{-0.6} - 0.96^{+0.37}_{-0.71} i $   \\
        \midrule 
         Exp. ( $K^{-} d $ )&~   $ a $ [fm] & ~$ r_{0} $ [fm] \\ 
        \midrule 
        SIDD-2 & $ ( 1.57 \pm 0.07 \pm 0.01)- (1.11  \pm 0.13 \pm 0.04)i$ & $-$   \\  
         ALICE & $ ( 1.44 \pm 0.15 \pm 0.10)- (1.34  \pm 0.33_{-0.15}^{+0.21} )i$ &  $-$ \\      
        \bottomrule 
    \end{tabular}
    \caption{Scattering length ($a$) and effective range ($r_0$) for $\olsi{K}^{0}d$ from the two theoretical models, compared with the $K^-d$ values from SIDDHARTA-2~\cite{Bazzi:2026gri} and ALICE~\cite{ALICE:2026pxr}.}
    \label{tab:aro}
\end{table*}

Using Eqs.~(\ref{eq:a0}) and~(\ref{eq:r0}), we evaluate $a$ and $r_0$ for the two models; results are given in Table~\ref{tab:aro}. The scattering lengths differ by only about $10\%$ between models and are compatible within their uncertainties. These $\olsi{K}^{0}d$ results should be contrasted with the $K^-d$ scattering lengths in the bottom panel of Table~\ref{tab:aro}, extracted by ALICE and SIDDHARTA-2 from the $K^-d$ CF and kaonic deuterium, respectively.\footnote{As discussed in Ref.~\cite{Bazzi:2026gri}, the SIDDHARTA-2 $K^-d$ scattering length is obtained from the measured $1s$ level shift and width via the summed-up Deser formula~\cite{Shevchenko:2021swf}.} The two determinations are consistent, with SIDDHARTA-2 reducing the uncertainty by more than a factor of $2$ and providing a less model-dependent result. The ALICE extraction instead relies on the Lednicky--Lyuboshitz formula~\cite{Lednicky:1981su}, whose limitations~\cite{Albaladejo:2024lam} suggest somewhat larger uncertainties than quoted in Table~\ref{tab:aro}.

The theoretical $\olsi{K}^{0}d$ scattering lengths are of the same order and share the same attractive character as the experimental $K^-d$ values, strong enough to generate a bound state, though somewhat smaller in magnitude -- naturally attributed to the additional effects of Coulomb attraction present in the $K^-d$ but not in the neutral $\olsi{K}^{0}d$ system.

The effective range, not yet estimated from experimental analysis, is more uncertain. Its real part is consistent between the two models. The imaginary part, however, differs substantially: the Oset--Ramos model~\cite{Oset:1997it}  yields a value compatible with zero, while the BCN model~\cite{Feijoo:2018den} yields a negative imaginary part about one third of the real part, although with large uncertainties. A future determination of $r_0$ would therefore help further constrain the $\olsi{K}^{0}d$ interaction.

\subsection{Correlation Function}
\begin{figure}
\centering
\includegraphics[width=\linewidth]{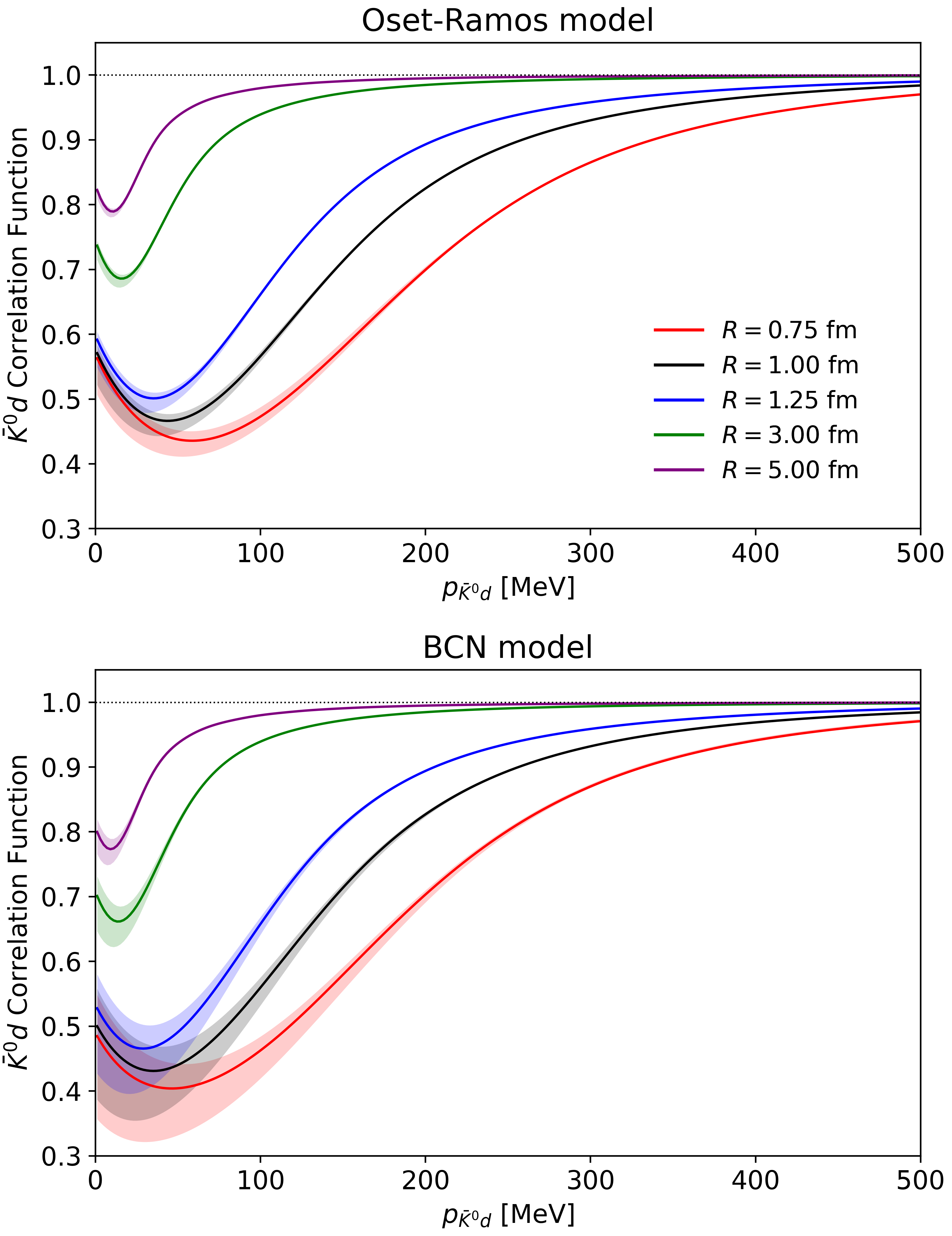}
\caption{Results for the $\olsi{K}^{0}d$ correlation function obtained with the Oset--Ramos~\cite{Oset:1997it} (upper plot) and BCN~\cite{Feijoo:2018den} (lower plot) models for different values of the source radius $R$.}
\label{fig:CorF}
\end{figure}
The $\olsi{K}^{0}d$ CF provides complementary information to that obtained from the recent measurement of the $K^-d$ CF by the ALICE Collaboration~\cite{ALICE:2026pxr}, and offers the possibility of further constraining the $\olsi{K}NN$ interaction and determining more precisely the properties of the predicted bound state. From an experimental point of view, the $K_Sd$ CF is the only experimentally accessible channel containing neutral kaons, since the $K_S$ can be efficiently reconstructed through its main decay into $\pi^+\pi^-$, while channels involving neutral pions cannot be reconstructed with the present ALICE setup. Since $\ket{K_S} =\left(\ket{K^0}+\ket{\olsi{K}^{0}}\right)/\sqrt{2}$, the $K_Sd$ CF receives contributions from both the $K^0d$ and $\olsi{K}^{0}d$ interactions. However, the $K^0N$ contribution can be safely neglected in the present analysis, as it is repulsive in the $I=1$ channel, vanishes at leading order in the $I=0$ channel, and is considerably weaker than the attractive $\olsi{K}N$ interaction~\cite{Oset:1997it}. Under this assumption, the $K_Sd$ CF is dominated by the $\olsi{K}^{0}d$ interaction. We therefore present the $\olsi{K}^{0}d$ CF which, within this approximation, determines the $K_S d$ CF as $C_{K_Sd} = (1 + C_{\olsi{K}^0d})/2$. 

An important feature of the present system is the absence of Coulomb interaction, which provides direct access to the low-energy scattering properties without the additional distortions present in charged systems. In particular, the shape and strength of the CF at small relative momenta are determined solely by the strong interaction, making the $\olsi{K}^{0}d$ CF a particularly clean observable for probing the low-energy interaction and, potentially, the properties of the predicted bound state.

Pending the corresponding experimental measurement, the $\olsi{K}^{0}d$ ($K_Sd$) CF is evaluated using the two theoretical models for several values of the source radius $R$, with the results shown in Fig.~\ref{fig:CorF}. The CFs obtained from the two models are compatible with each other within their respective uncertainties. Moreover, the differences between their central values decrease as the source radius increases, as expected. These observations illustrate the difficulty of discriminating between the models through a potential experimental measurement of such a CF given the current experimental uncertainties \cite{ALICE:2026pxr}, especially at low momenta. This limitation is not specific to the models considered here, but is expected to extend to a broad class of chiral models, which are generally constrained by threshold observables and low-energy scattering data and tend to differ more substantially at $\bar K N$ subthreshold energies, as illustrated, for example, in Fig.~1 of Ref.~\cite{Cieply:2020ftt}. As experimental precision improves with future data, the uncertainties of the measured $\olsi{K}^{0}d$ ($K_Sd$) CF are expected to decrease accordingly, allowing for a more sensitive comparison between the theoretical models.

Nevertheless, a characteristic CF shape associated with a system containing a bound state below threshold is found, similar to that observed previously for the $pf_1(1285)$ interaction and related reactions. In case such CF is measured with similar shape, it would confirm the existence of the state found by the theoretical models.  

\section{Conclusions}
\label{sec:conclusions}

We have carried out calculations for the scattering amplitude and CF of the $\olsi{K}^0d$ system. This system is closely related to the $K^-d$ and $K^-pp$ systems, which have received much attention, but has the advantage, when it comes to the CF, of being free from the Coulomb interaction. This allows the data to be interpreted more cleanly in terms of the strong interaction alone, without the interference between strong and Coulomb effects present in the charged channels.

The $\olsi{K}^0d$ scattering amplitude has been obtained using a reformulation of the FCA to the Faddeev equations that implements exact elastic unitarity, a feature that is important for a reliable evaluation of the low-energy scattering parameters, $a$ and $r_0$, and of the CF.
Below threshold, we find a clean resonance structure about $40$--$60$~MeV below the $\olsi{K}^0d$ threshold, with a width of about $50$--$75$~MeV. The position of the resonance corresponds to the formation of a bound state between a nucleon and the lower pole of the $\Lambda(1405)$, around $1380$~MeV, thus providing a picture complementary to the $K^-pp$ bound state, which is instead associated with the higher $\Lambda^*$ pole. To estimate the theoretical uncertainty, all observables have been evaluated with two different $\olsi{K}N$ interaction models. The resonance position differs by about $20$~MeV between the two models, while the scattering length differs by only about $10\%$, indicating that the low-energy scattering parameters are considerably more robust against the choice of underlying interaction than the detailed position and width of the bound state itself.

Although the $\olsi{K}^0d$ and $K^-d$ scattering amplitudes need not coincide, the values obtained here are of the same order of magnitude as those extracted from recent experiments, the residual difference being naturally attributed to the Coulomb interaction present only in the charged system. We also argue that the bound state found here is related to the already observed $K^-pp$ bound state, making a dedicated search for this new state particularly interesting.
Measurement of the $\olsi{K}^0d$ CF should be feasible for the ALICE Collaboration through the experimentally accessible $K_Sd$ channel, which is expected to be dominated by the $\olsi{K}^0d$ interaction, and which the same Collaboration has already measured and analyzed for the charged $K^-d$ case. A complementary probe could also come from photoproduction reactions such as $\gamma d\to K^0\Sigma^0p$, where structures in the $M_{\Sigma p}$ invariant-mass distribution could provide additional access to the predicted bound state. The predictions presented here should provide an incentive for new experiments that would complement the information already obtained from the $K^-d$ and $K^-pp$ systems.

\bibliography{K0d.bib}% Produces the bibliography via BibTeX.

\end{document}